\documentclass[aps,prl,twocolumn,showpacs]{revtex4}

\usepackage{dcolumn}                    
\usepackage{bm}                         
\usepackage{graphicx}

\newcommand {\barr} {\begin{eqnarray}}
\newcommand {\earr} {\end{eqnarray}}
\newcommand {\beq} {\begin{equation}}
\newcommand {\eeq} {\end{equation}}

\usepackage{times}

\begin{document}

\title{Volovik effect on NMR measurements of unconventional superconductors}

\author{Yunkyu Bang}
\email[]{ykbang@chonnam.ac.kr} \affiliation{Department of Physics,
Chonnam National University, Kwangju 500-757, Republic of Korea}
\begin{abstract}
We studied the Volovik effect on the NMR measurements of the two
unconventional superconducting (SC) states, the d-wave and
$\pm$s-wave states. We showed the generic field dependencies of
the spin-lattice relaxation rate $1/T_1$ and Knight shift $K$ at
low temperature limit in the pure cases as: $1/T_1 \propto H \log
H$, $K \propto \sqrt{H}$ for the d-wave, and $1/T_1 \propto H$, $K
\propto H$ for the $\pm$s-wave state, respectively. Performing
numerical calculations we showed that these generic power laws
survive for the good part of low field region with the realistic
amount of impurities.
We also found that the Volovik effect acts as an equivalent pair
breaker as the unitary impurity scattering, hence induces the same
temperature evolutions on $1/T_1$ and $K$, respectively, as the
unitary impurities in both SC states. This finding implies that
the Volovik effect should always be taken into account for the
analysis of the NMR measurements in the mixed state.
\end{abstract}


\date{\today}
\maketitle

{\it Introduction. --} There are various experimental probes of
the gap symmetry of the superconductors. Mostly, they are probing
low lying excitations in the superconducting (SC) state to
distinguish a full gap, a nodal, or a point gap. The measurement
results usually produce the typical temperature dependencies, for
example, an exponentially flat for a full gap or various power
laws for a nodal or point gap.
The Volovik effect (VE) \cite{Volovik} has enlarged the probing
dimension to the magnetic field axis. The principle is actually
very simple in that the low energy density of states (DOS) of the
SC state is changed in the mixed state with vortices induced by
the applied magnetic fields and then its field dependence is
probed. In particular, this change of DOS due to the vortices
should be more sensitive with a gapless superconductor such as the
d-wave state in the cuprate superconductors \cite{Volovik,
Kubert}.
However, recently, it was shown that the strong field dependence
of the Volovik effect is not unique to the nodal gap
superconductors but also can appear with the full gap s-wave
superconductors if there exist multiple s-wave gaps with different
gap sizes \cite{Bang Volovik}.


On the other hand, most of the Volovik effect measurements have
been carried out with the specific heat (SH) and thermal
conductivity. However, in principle, it should be effective with
any experimental measurements which probe the low energy DOS and
its variation with the applied field. Therefore, the NMR
measurements such as the spin-lattice relaxation rate $1/T_1
(T,H)$ and Knight shift $K(T,H)$ should also be good probes for
the Volovik effect. In comparison with the SH and thermal
conductivity measurements, the NMR measurements of Volovik effect
require much harsher conditions such as the sample size, low
temperature control, and high quality homogeneity of the field,
etc.
Nevertheless there already exist a few experimental works, for
example, on Fe pnictide superconductors,
Ba$_{0.69}$K$_{0.31}$Fe$_2$As$_2$ by W. P. Halperin and
coworkers\cite{Halperin} and
BaFe$_2$As$_2$(As$_{0.67}$P$_{0.33}$)$_2$ by Nakai {\it et.
al.}\cite{Nakai}, as well as on the high-$T_c$ cuprates by G. Q.
Zheng and coworkers \cite{GQ Zheng cuprates}. However, because the
theoretical study on this subject is very rare \cite{Vekhter}, the
reliable interpretation and the extraction of the useful
information from the experimental data are limited. Therefore more
systematic and detailed theoretical study of the Volovik effects
on the NMR measurements is highly demanded and it will enhance the
capability of the NMR technique to study the superconducting gap
symmetry.

In this paper, we specifically studied the NMR $1/T_1$ and Knight
shift $K$ in the SC state of the two typical unconventional
superconductors, namely, the d-wave and $\pm$s-wave
states\cite{Mazin, Bang model}. We derived the exact power laws of
the generic field dependencies of both SC cases in pure state:
$1/T_1 \propto H \log H$, $K \propto \sqrt{H}$ for the d-wave, and
$1/T_1 \propto H$, $K \propto H$ for the $\pm$s-wave state,
respectively.
We also numerically studied the impurity effects on $1/T_1$ and
$K$, using the self-consistent $T$-matrix approximation (SCTA)
\cite{T-mtx, Bang-imp}, and provided the systematic comparison
between the d-wave and $\pm$s-wave gap states.
We found that the impurity scattering substantially reduces the
low field region where the generic field dependencies survive in
the $\pm$s-wave case, while it doesn't much affect the generic
power laws except a constant shifting in the d-wave case
\cite{note1}.
Another key finding of our numerical study is that the Volovik
effect induced by magnetic fields practically acts as an
equivalent pair breaker as the strong coupling (unitary limit)
impurities, hence modifies the low temperature behaviors of the
$1/T_1$ and $K$ in the same fashion as due to the formation of the
resonant impurity band. This result implies that it is always
necessary to take the Volovik effect into account when one
interprets the low temperature NMR experiments with magnetic field
in order to extract the correct information from the data.

{\it Formalism. --} For most of purposes to study the Volovik
effect, we just need to calculate the position dependent DOS
$N(\omega, r)$ in the presence of vortices. Using the
semiclassical approximation, the matrix form of the
single-particle Green's function in the SC state, including
Doppler shift of the quasiparticle excitations $\epsilon (k)$ due
to the circulating supercurrent ${\bf v_s (r)}$, is given by
\cite{Volovik,Kubert}

\beq
\hat{G} ({\bf k, r,\omega})=\frac{[\omega +  {\bf v_s (r)} \cdot
{\bf k}] \tau_0 + \epsilon (k) \tau_3 + \Delta \tau_1}{[ \omega +
{\bf v_s (r)} \cdot {\bf k}]^2 - \epsilon ^2 (k) - \Delta ^2 }
\eeq

\noindent where $\tau_i$ are Pauli matrices and ${\bf r}$ is the
distance from the vortex core. $\Delta$ is the SC gap function and
${\bf v_s (r)}$ is  $\sim \frac{1}{m} \frac{1}{r} {\bf
\hat{\theta}}$. The position dependent DOS is calculated as $N
(\omega,r)= - \frac{1}{\pi} {\rm Tr Im} \sum_k G_{0} ({\bf k,
r,\omega})$. Finally, the field dependent quantities are obtained
from the areal average DOS per unit volume as $\bar{N}
(\omega,H)=\int_{\xi} ^{R_H} dr^2 N (\omega,r) / \pi R_H ^2$ with
the magnetic length $R_H = \sqrt{\frac{\Phi_0}{\pi H}}$ ($\Phi_0$
a flux quanta) and the SC coherence length $\xi$.

In the homogenous SC state, the general structure of the $1/T_1$
is written as

\beq \frac{1}{T_1} =  - T \int_0 ^{\infty} d\omega \frac{\partial
f_{FD} (\omega)}{\partial \omega} \left[
 N(\omega)^2 +  M(\omega)^2
\right], \eeq
and the Knight shift $K$ as  \beq K = - T \int_0 ^{\infty} d\omega
\frac{\partial f_{FD} (\omega)}{\partial \omega} N(\omega), \eeq

\noindent where  $N(\omega)=
 \langle Re\frac{\omega}{\sqrt{\omega^2-\Delta^2(\theta)}}
 \rangle_{\theta}$ is the Fermi surface (FS) averaged DOS and $
M(\omega)= \langle Re
\frac{\Delta(\theta)}{\sqrt{\omega^2-\Delta^2(\theta)}}
\rangle_{\theta}$ is the similar quantity induced in the SC state.
In the mixed state, we only need to replace $N(\omega)$ by
$\bar{N}(\omega, H)$ and $M(\omega)$ by $\bar{M}(\omega, H)$ in
the above formulas (2) and (3). $\bar{M}(\omega, H)$ is obtained
from $M (\omega,r)= - \frac{1}{\pi} {\rm Tr Im} \sum_k G_{1} ({\bf
k, r,\omega})$ and its areal average. Then we can calculate the
field as well as temperature dependent NMR properties $1/T_1
(T,H)$ and $K(T, H)$.

In the mixed state of the d-wave superconductor, it is well known
that the $N (\omega=0,r) \sim 1/r$ due to the linear DOS $N
(\omega) \sim \omega$ and the Doppler shifting energy ${\bf v_s
(r)} \cdot {\bf k_F} \sim 1/r$. Also knowing that $M(\omega)$
vanishes due to the gap symmetry, we can readily extract the field
dependence of the $1/T_1$ for the d-wave state from Eq.(2) as

\beq  \frac{1}{T_1}(T \rightarrow 0, H) \sim \int_{\xi} ^{R_H}
dr^2 (\frac{1}{r})^2 / \pi R_H ^2 \sim H \log H, \eeq

\noindent and similarly for Knight shift as

\beq K(T \rightarrow 0, H) \sim \int_{\xi} ^{R_H} dr^2
(\frac{1}{r}) / \pi R_H ^2 \sim \sqrt{H}. \eeq

In the case of the $\pm$s-wave state with small gap $\Delta_S$ and
large gap $\Delta_L$, again all we need to know is the DOS near
zero energy in the mixed state. The author\cite{Bang Volovik} has
recently shown that the Doppler shifting energy ${\bf v_s (r)}
\cdot {\bf k_F} \sim \Delta_L/r$ overshoots the small gap to
induce a constant DOS -- which is basically the normal state DOS
of the small gap band $N_S^{norm}$-- around $\omega=0$ in the
small gap band in the region around the vortex core to the
distance $r^* = \xi \frac{\Delta_L}{\Delta_S}$ that is a field
independent constant. Hence the field dependencies of the NMR
properties of the $\pm$s-wave state are the following.

\beq  \frac{1}{T_1} (T \rightarrow 0, H) \sim \int_{\xi} ^{r^*}
dr^2 (N_S^{norm})^2 / \pi R_H ^2 \sim H  \eeq

and

\beq K(T \rightarrow 0, H) \sim \int_{\xi} ^{r^*} dr^2
(N_S^{norm}) / \pi R_H ^2 \sim H \eeq

\noindent where the contribution of $M(\omega)$ term is again
neglected since its contribution is significant only near $T_c$
and becomes negligible at low temperatures. The above field
dependencies are for the ideally pure cases and we will show
below, with numerical calculations, how these generic field
dependencies are modified when the impurity scattering effect is
included.

{\it Numerical results and discussions.--} We calculate the
$1/T_1(T,H)$ and $K(T,H)$ including the impurity scattering effect
using the SCTA\cite{T-mtx, Bang-imp}. Once we calculate the
impurity induced selfenergy -- normal and anomalous --
corrections, $\Sigma_{imp}^{0}(\omega, r)$ and
$\Sigma_{imp}^{1}(\omega, r)$, respectively, we renormalize
everywhere in the above formalism (Eqs. (1) -(3)) by
$\tilde{\omega}=\omega+\Sigma_{imp}^{0}(\omega, r)$ and
$\tilde{\Delta}=\Delta+\Sigma_{imp}^{1}(\omega, r)$
($\Sigma_{imp}^{1}$ vanishes in the d-wave case due to the gap
symmetry). In this paper, we considered only the strong coupling
(unitary limit) non-magnetic impurity. The extension to the
$\pm$s-wave state for the calculations of $N(\omega, r)$ and
$M(\omega, r)$ with two bands is straightforward and referred to
Ref.\cite{Bang Volovik}.

Figures 1 and 2 are the calculation results of the d-wave SC
state. Fig. 1(a) shows the normalized $1/T_1(T)$ of the pure
d-wave state with various field strengths. First, the zero field
data displays an exact $T^3$ power law for all temperatures below
$T_c$ (for this we chose $2\Delta_0/T_c =4$). Applying fields, the
low temperature behavior immediately changes to the $T$-linear
power indicating that the zero energy DOS is created by the
magnetic field. Increasing the field strength, the magnitude of
$1/T_1(T)$ at low temperatures progressively increases. In the
inset, we plotted $1/T_1(T=0.05T_c)$ vs $H/H_c$. As predicted in
Eq.(4), the numerical data points (red circles) are perfectly
fitted with $H \log{H}$ curve (black solid line; the fitting
function is given in the inset with $h=H/H_{c2}$).

\begin{figure}
\hspace{0cm}
\includegraphics[width=90mm]{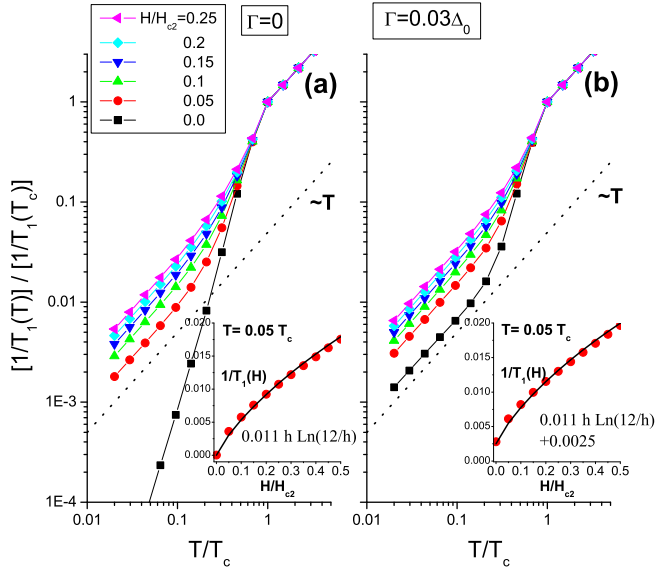}
\caption{(Color online) Normalized spin-lattice relaxation rates
$[1/T_1(T) / 1/T_1(T_c)]$ of the d-wave SC state with
$2\Delta_0/T_c =4$ for various magnetic field strengths,
$H/H_{c2}=0.0, 0.05, 0.1, 0.15, 0.2$ and 0.25, respectively. (a)
The results without impurities $(\Gamma=0)$. (b) The results with
the impurity concentration $\Gamma=0.03 \Delta_0$ of the unitary
scatterer. Insets of each panel are the $1/T_1$ vs $H/H_{c2}$ at a
fixed low temperature $T=0.05 T_c$.} \label{fig.1}
\end{figure}

\begin{figure}
\hspace{0cm}
\includegraphics[width=90mm]{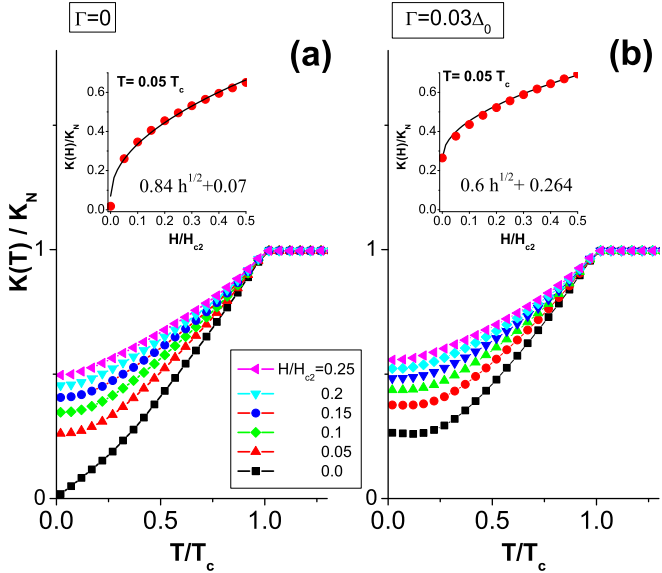}
\caption{(Color online) Normalized Knight shift $K(T) / K(T_c)$ of
the d-wave SC state for various magnetic fields strengths,
$H/H_{c2}=0.0, 0.05, 0.1, 0.15, 0.2$ and 0.25, respectively. (a)
The results without impurities $(\Gamma=0)$. (b) The results with
the impurity concentration $\Gamma=0.03 \Delta_0$ of the unitary
scatterer. Insets of each panel are the $K(H)/K_N$ vs $H/H_{c2}$
at a fixed low temperature $T=0.05 T_c$.} \label{fig.2}
\end{figure}

Figure 1(b) shows the same calculations as in Fig.1(a) but with
impurity scattering. We chose the impurity concentration parameter
$\Gamma= n_{imp}/\pi N_{tot}(0) = 0.03 \Delta_0$ of the
non-magnetic unitary scatterers. At zero field (black squares),
$1/T_1(T)$ shows the typical $T$-linear at low temperatures due to
the resonant (zero energy) impurity band but maintains the $T^3$
power law at higher temperatures. Applying fields, the overall
line shape doesn't change but only the low temperature $T$-linear
part progressively moves upward. It is clear that the zero energy
DOS induced by the impurity scattering is additive to the field
induced zero energy DOS by the Volovik effect. This result was not
{\it a priori} expected because we calculated the position
dependent impurity selfenergy $\Sigma_{imp}^{0} (\omega, r)$
treating the Volovik effect and the impurity scattering on equal
footing before we take the areal average. The plot of
$1/T_1(T=0.05T_c)$ vs $H/H_c$ in the inset starts with a finite
intercept at zero field, but still the data points are well fitted
with $a H \log{H} + const.$ (black solid line).

Figure 2 is the calculation results of the normalized Knight shift
$K(T)$ in parallel with the $1/T_1(T)$ in Fig.1. It reflects the
same evolution of the DOS $\bar{N}(\omega, H)$ of the d-wave state
with magnetic fields and impurities, as explained in Fig.1, onto
$K(T)$. Fig.2(a) shows the normalized Knight shift $K(T)$ of the
pure d-wave gap with various field strengths. First, the zero
field data displays the $T$-linear power law at low temperatures
as expected for the nodal gap. Applying fields, the low
temperature part immediately becomes flat because of the field
induced zero energy DOS and progressively moves upward with
increasing the field strength. The inset shows $K(T=0.05T_c)$ vs
$H/H_c$ data and well fitted by the $\sqrt{H}$ curve (black solid
line) as shown in Eq.(5).
Fig.2(b) shows the same calculations of $K(T)$ with impurity
scattering as in Fig.1(b). The overall behavior and evolution of
it are now easily understood with the additive zero energy DOSs
induced by impurities and by the magnetic field. The plot of
$K(T=0.05T_c)$ vs $H/H_c$ data in the inset is well fitted by the
$a \sqrt{H} + const.$ curve \cite{note1}.

Figures 3 and 4 are the calculation results of $1/T_1$ and $K$ of
the $\pm$s-wave SC state. We assumed two gaps with different
sizes, $\Delta_{S} / \Delta_{L}=0.4$. Other parameters are: (1) A
rather large value of $2\Delta_{L}/T_c =10$ was chosen to produce
the much steep drop of $1/T_1(T)$ below $T_c$; (2) $N_L (0) /N_S
(0) =2$ is freely chosen although there is a correlation between
two ratios, $\Delta_{S} / \Delta_{L}$ and $N_L (0) /N_S (0)$
\cite{Bang model}. As in the case of d-wave calculations, we
considered only the unitary scatterers and assumed an equal
strength between the interband and intraband scattering channels
\cite{Bang-imp}.

Figure 3(a) shows the normalized $1/T_1(T)$ of the pure
$\pm$s-wave state for various field strengths. First, the zero
field $1/T_1(T)$ (black squares) displays the exponential drop for
all temperature range without the coherence peak. The initial drop
can be fitted by the $T^6$ power, much steeper --due to the choice
of $2\Delta_{L}/T_c =10$ --  than the $T^3$ power that was often
observed with many of Fe pnictide SC compounds. Applying fields,
the low temperature behavior immediately changes to the quasi
$T$-linear because of the zero energy DOS induced by the vortices
as in the case of the d-wave gap. Increasing the field strength,
the magnitude of $1/T_1(T)$ at low temperatures progressively
increases. In the inset, we plotted $1/T_1(T=0.05T_c)$ vs $H/H_c$
and the numerical data points (red circles) are perfectly fitted
with the $H$-linear line (black dotted line) before it saturates
when the small gap band becomes completely collapsed beyond the
critical field strength $H^{*}/H_{c2} \sim (\Delta_{S} /
\Delta_{L})^2$ as predicted in Ref.\cite{Bang Volovik}.

Figure 3(b) shows the same calculations of $1/T_1(T)$ but with
impurity scattering. The value of the impurity concentration
$\Gamma=0.08 \Delta_L$ was slightly larger than the critical
concentration of the $\pm$s-wave model \cite{Bang-imp}
($\Gamma^{*} \approx 0.045 \Delta_L$ in this case), which would
produce a perfect V-shape DOS and hence the $T^3$ power law at low
temperatures. Therefore, in our case, the zero field data of
$1/T_1(T)$ (black squares) shows some power law in between $T^3$
and $T$-linear at low temperatures due to the combination of the
finite zero energy DOS plus the V-shape DOS.
Applying fields, the low temperature behavior immediately changes
to the quasi $T$-linear as in the pure case of Fig.3(a) and the
overall magnitude progressively increases with increasing the
field strength. The plot of $1/T_1(T=0.05T_c)$ vs $H/H_c$ in the
inset of Fig.3(b) is much smoothened and there is no sharp kink at
the saturation field $H^{*}\approx 0.16 H_{c2}$; but the
$H$-linear behavior still survives in the reduced region of low
fields.

\begin{figure}
\hspace{0cm}
\includegraphics[width=90mm]{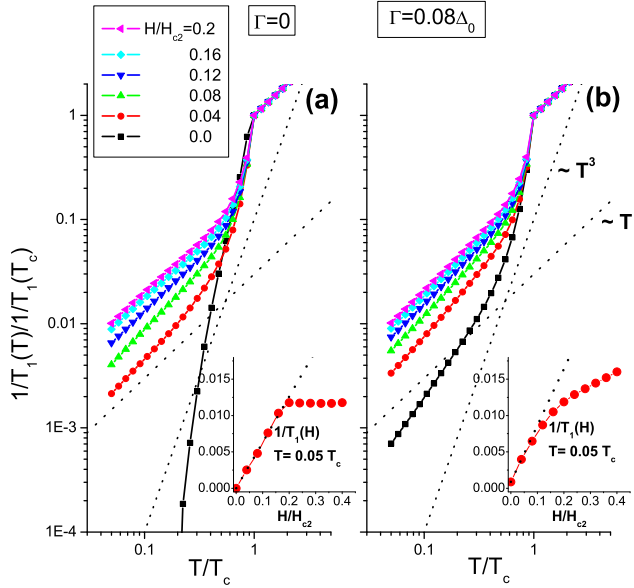}
\caption{(Color online) Normalized spin-lattice relaxation rates
$[1/T_1(T) / 1/T_1(T_c)]$ of the $\pm s$-wave SC state for various
magnetic fields $H/H_{c2}=0.0, 0.04, 0.08, 0.12, 0.16$ and 0.2,
respectively. (a) The results without impurities $(\Gamma=0)$. (b)
The results with the impurity concentration $\Gamma=0.08 \Delta_L$
of the unitary scatterer. Insets of each panel are the $1/T_1$ vs
$H/H_{c2}$ at a fixed low temperature $T=0.05 T_c$. Other
parameters: $2\Delta_L/T_c =10$, $\Delta_S / \Delta_L =0.4$.}
\label{fig.3}
\end{figure}

\begin{figure}
\hspace{0cm}
\includegraphics[width=90mm]{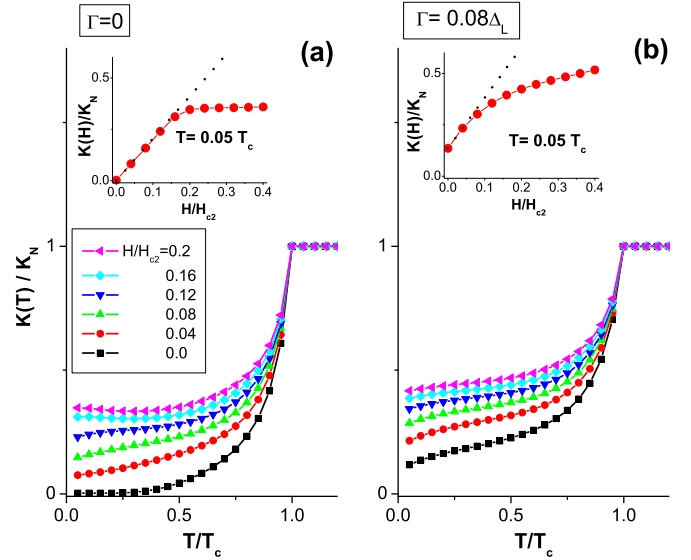}
\caption{(Color online) Normalized Knight shift $K(T) / K(T_c)$ of
the $\pm s$-wave SC state for various magnetic fields
$H/H_{c2}=0.0, 0.04, 0.08, 0.12, 0.16$ and 0.2, respectively.  (a)
The results without impurities $(\Gamma=0)$. (b) The results with
the impurity concentration $\Gamma=0.08 \Delta_L$ of the unitary
scatterer. Insets of each panel are the $K(H)/K_N$ vs $H/H_{c2}$
at a fixed low temperature $T=0.05 T_c$.} \label{fig.4}
\end{figure}

Figure 4 is the calculation results of the normalized Knight shift
$K(T)$ of the $\pm$s-wave SC state. Fig.4(a) shows the $K(T)$ of
the pure $\pm$s-wave gap with various field strengths. First, the
zero field data displays the exponentially flat behavior at low
temperatures reflecting the full gap superconductor.
Applying fields, the low temperature part moves upward and also
develops a $T$-linear part up to $H/H_{c2} \approx 0.12$. This
behavior of $K(T) \sim a+b T$ at low temperatures is the result of
the combination of the finite zero energy DOS plus the V-shape DOS
-- this is the typical feature of the multi-band $\pm$s-wave SC
state with the resonant impurity scattering\cite{Bang-imp} but it
is now generated by the magnetic fields. The inset shows
$K(T=0.05T_c)$ vs $H/H_c$ data which is well fitted by the
$H$-linear line (black dotted line) as shown in Eq.(7) at low
fields.

Figure 4(b) shows the same calculations of $K(T)$ including
impurity scattering of $\Gamma=0.08 \Delta_L$. The overall
behavior and evolution of it are again easily understood with the
addition of the impurity induced DOS and the magnetic field
induced DOS. The inset plot of $K(T=0.05T_c)$ vs $H/H_c$ data is
much rounded and no kink feature is visible as in the inset of
Fig.3(b). In general, all data of $K(T)$ calculations in Fig.2 and
Fig.4 faithfully reflect the evolution of the total DOS
$N_{tot}(\omega)$ at low frequencies by magnetic fields and
impurity scattering as can be seen from Eq.(3).

{\it Conclusion. --} In summary, we extended the study of the
Volovik effect on the NMR properties both in the d-wave and
$\pm$s-wave gap states. We derived the generic power laws of the
field dependencies of $1/T_1(T \rightarrow 0, H)$ and $K(T
\rightarrow 0, H)$ in pure cases. We also numerically calculated
the full temperature dependencies of $1/T_1(T,H)$ and $K(T,H)$ for
various field strengths, for both SC states, including impurity
scattering. Main findings are: (1) the field dependence power laws
of $1/T_1(T \rightarrow 0, H)$ and $K(T \rightarrow 0, H)$ are
practically unchanged with impurity scattering in the d-wave case
except for a constant shifting\cite{note1}. In the $\pm$s-wave
case, while the changes of the overall field dependencies are more
significant, the generic power laws survive in a good part of low
field region; (2) the Volovik effect acts as an equivalent pair
breaker as the unitary impurity scatterers to creates the zero
energy DOS and hence to modify the low temperature power laws of
$1/T_1(T)$ and $K(T)$. It implies that the Volovik effect always
needs to be taken into account for the analysis of the NMR
experiments.

{\it Acknowledgement -- } The author are grateful to Professor W.
P. Halperin for showing us their data prior to publication and
intensive discussions which directly motivated this work. We also
acknowledge the discussions with G.Q. Zheng. The author (Y.B.) was
supported by the Grant No. NRF-2010-0009523 and NRF-2011-0017079
funded by the National Research Foundation of Korea.


\begin{references}

\bibitem{Volovik}
G. E. Volovik, JETP Lett. {\bf 58}, 469 (1993).

\bibitem{Kubert}
C. Kubert and P.J. Hirschfeld, Solid State Commun. {\bf 105}, 459
(1998).

\bibitem{Bang Volovik}
Y. Bang, Phys. Rev. Lett. 104, 217001 (2010).


\bibitem{Halperin}
S. Oh, A. M. Mounce, W. P. Halperin, C. L. Zhang, P. Dai, A. P.
Reyes, P. L. Kuhns, arXiv:1109.3834 (unpublished).

\bibitem{Nakai}
Y. Nakai, T. Iye, S. Kitagawa, K. Ishida, S. Kasahara, T.
Shibauchi, Y. Matsuda, and T. Terashima, Phys. Rev. B {\bf 81},
020503 (2010).

\bibitem{GQ Zheng cuprates}
S. Kawasaki, C. Lin, P. L. Kuhns, A. P. Reyes, and Guo-qing Zheng,
Phys. Rev. Lett. {\bf 105}, 137002 (2010); Guo-qing Zheng, H.
Ozaki, Y. Kitaoka, P. Kuhns, A. P.Reyes, and W. G. Moulton, Phys.
Rev. Lett. {\bf 88}, 077003 (2001).

\bibitem{Vekhter}
I. Vekhter, P. J. Hirschfeld, and E. J. Nicol, Phys. Rev. B, {\bf
64}, 064513 (2001).

\bibitem{Mazin}
I.I. Mazin, D.J. Singh, M.D. Johannes, M.H. Du, Phys. Rev. Lett.
{\bf 101}, 057003 (2008); K. Kuroki, S. Onari, R. Arita, H. Usui,
Y. Tanaka, H. Kontani, and H. Aoki , Phys. Rev. Lett. {\bf 101},
087004 (2008).

\bibitem{Bang model}
Y. Bang and H.-Y. Choi, Phys. Rev. B, {\bf 78}, 134523 (2008).


\bibitem{T-mtx}
P. J. Hirschfeld, P. Wolfle, and D. Einzel, Phys. Rev. B {\bf 37},
83 (1988); A. V. Balatsky, I. Vekhter, and J.-X. Zhu, Rev. Mod.
Phys. {\bf 78}, 373 (2006).

\bibitem{Bang-imp}
Y. Bang, H.-Y. Choi, and H. Won, Phys. Rev. B {\bf 79}, 054529
(2009).

\bibitem{note1}
The dirty limit region where $\sqrt{H}$ behavior is replaced by
the $H \ln{H}$ form is restricted to the very low field region,
$H/H_{c2} \ll 0.36~ \Gamma/\Delta_0$ \cite{Kubert}, which is
$H/H_{c2} \ll 0.01$ in our calculations with $\Gamma/\Delta_0
=0.03$.



\end{references}
\end{document}